# RSMS: Towards Reliable and Secure Metaverse Service Provision

Yanwei Gong, Xiaolin Chang, Jelena Mišić, Vojislav B. Mišić, Yingying Yao

*Abstract*—Establishing and sustaining Metaverse service necessitates an unprecedented scale of resources. This paper considers the deployment of Metaverse service in a cloud-edge resource architecture, which can satisfy the escalating demand for Metaverse service resources while ensuring both high bandwidth and low latency.

We propose a novel mechanism, named *R*eliable and *S*ecure *M*etaverse *Ser*vice (RSMS), to ensure Metaverse service reliability and security without sacrificing performance. RSMS consists of two protocols: (1) One is a blockchain-based lightweight mutual authentication protocol concerning heterogeneous Metaverse service resource nodes (RNs) dynamically joining a Metaverse service resource pool while guaranteeing their trustworthiness, which guarantees the security of Metaverse service. (2) The other is a group authentication protocol used to form and maintain a stable and secure Metaverse service group composed by RNs, which ensures the reliability and enhances the security of Metaverse service. The reliability and security of Metaverse service under RSMS are thoroughly discussed, and informal and formal security analysis are conducted. Additionally, we study the impact of RSMS on Metaverse service throughput, demonstrating its lightweight feature.

*Index Terms*—Cloud-edge, Group Authentication, Lightweight, Metaverse service, Reliability

## I. INTRODUCTION

The concept of Metaverse, which was initially introduced in 1992 [1], has garnered increasing interest from both academia and industry in recent years. It can be attributed to the notable advancements in various technologies, including extended reality, 5G/6G networks, and edge intelligence. Additionally, major corporations like Facebook [2] and Microsoft [3] have made significant contributions, further propelling Metaverse into the spotlight. Metaverse is expected to bring a new revolution to the digital world [4]. Nonetheless, one of the most formidable obstacles in the path of Metaverse deployment is the requirement of the immense resources it demands [4]. This challenge has hindered its widespread adoption. Although only utilizing conventional cloud computing infrastructure for Metaverse deployment is a potential solution, it introduces its own set of difficulties, including issues related to congestion and high latency [5].

A promising solution for the above challenges is a cloud-edge resource architecture, where the computing, storage, networking, and communication capabilities are distributed along the path from end users to cloud resources [6][7][8]. Fig.1 illustrates the Metaverse service system framework considered in this paper, which is also studied in [9]. The lower layer is Metaverse service resource nodes (RNs), which have available computing and storage resources. Each RN can be a virtual machine, container, physical machine in the cloud or at the network edge. They have different computing and storage resources, and different security levels. The resources possessed by RNs are essential to ensure the Metaverse services provision. Each RN belongs to a Metaverse service resource pool, managed by its Metaverse service resource pool manager (RPM). All RPMs consists of the middle layer. The top layer is Metaverse service resource provider (MSRP), responsible for receiving Metaverse service resource requests from users and allocates Metaverse service resources in the form of RNs. In Fig.1, there are three RPMs (A, B and C) and a MSRP. For each Metaverse service request from users, MSRP allocates a Metaverse service group of RNs from any pool. All the RNs in a Metaverse service group collaboratively work for the user.

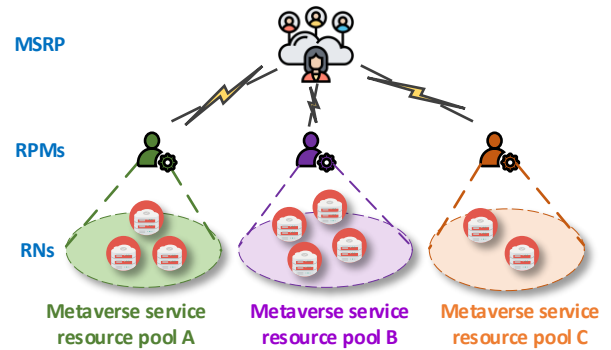

Fig.1. The illustration of Metaverse service system framework

In this paper, we focus on reliable and secure Metaverse service provision. In terms of reliability, we focus on the service continuity even some serving RNs can not work due to unexpected events like attacks and software aging. In terms of security, we mainly focus on confidentiality, integrity, and non-repudiation of transmitted messages during the Metaverse service provision. All of them are explained in detail in Section III.D. There are at least three key challenges to be addressed.

- Trustworthiness of RN. Each time an RN joins a Metaverse service resource pool, it must be authenticated to ensure its trustworthiness. Besides, each RN is dynamic and may choose to join different Metaverse service resource pools to obtain more profits. Therefore, authenticating RNs across Metaverse service resource pools must also be satisfied.
- Reliability of Metaverse service. During Metaverse service provision, an RN may not work well, affecting the reliability and security of Metaverse service. Then new RNs should be assigned to the Metaverse service group.
- Lightweight. Considering the huge number of RNs authentication and the necessity of RN authentication across Metaverse service resource pools, which may consume a

Y.Gong, X. Chang and Y. Yao are with Beijing Key Laboratory of Security and Privacy in Intelligent Transportation, Beijing Jiaotong University, Beijing, 100044, China. E-mails: {22110136, xlchang, yyyao}@bjtu.edu.cn. (Corresponding author: Xiaolin Chang.)
J. Mišić is with Ryerson University, Canada. Email: jmisic@scs.ryerson.ca.
V. B. Mišić is with Toronto Metropolitan University, Canada. Email: vmisic@torontomu.ca.

great deal of resources, the authentication protocol must be lightweight.

To achieve reliable and secure Metaverse service provision, we propose a novel mechanism, named *R*eliable and *S*ecure *M*etaverse *S*ervice (RSMS), which consists of a mutual authentication protocol and a group authentication protocol. The former is used for establishing secure Metaverse service resource pools. The latter aims for forming and maintaining the Metaverse service group composed of RNs to guarantee reliable and secure Metaverse service.

This paper assumes the MSRP has deployed defense mechanisms for detecting abnormal behaviors of RNs. Any malicious behavior of RNs can be detected and then its joining a Metaverse service resource pool will be revoked. Besides, when an RN breaks down during the provision of services, one or more new RNs will be allocated to replace it to ensure Metaverse service's reliability. Note that, how to monitor RNs behaviors, detect abnormality, and allocate RNs are all beyond the scope of this paper. There are some related works on malicious behavior detection [10][11] and resource allocation [12][13] about Metaverse service for reference.

To the best of our knowledge, this is the first proposal to explore a mechanism for provisioning reliable and secure Metaverse service. Our contributions are as follows:

- We design a blockchain-based lightweight mutual authentication protocol to ensure the trustworthiness of RNs, which will join a Metaverse service resource pool. This protocol also supports RN authentication across Metaverse service resource pools. We use a bloom filter and lightweight cryptographic primitive to reduce the computation and communication overhead. The protocol satisfies the security attributes of confidentiality, integrity, non-repudiation, anonymity, traceability, and unlinkability.
- We propose a group authentication protocol for authenticating RNs assigned to the user to provide Metaverse service resources by using secret sharing and elliptic curve cryptography. The group membership authentication and group session key agreement subprotocols of the group authentication protocol enhance the security of Metaverse service. Moreover, the group session key update subprotocol realizes RN seamless complementation, which satisfies forward and backward security of RNs exiting and joining the Metaverse service group and ensures the reliability of Metaverse service.
- We present a detailed analysis of reliability and security attributes, which includes informal and formal analysis. Besides, we also conduct Monte Carlo simulations to evaluate the efficiency of RSMS. Our simulation results indicate that, under various parameter settings, RSMS produces less effect on Metaverse service provision performance in terms of the number of completed Metaverse service requests in a period.

The remainder of the paper is arranged as follows. Related work is presented in Section II. The system framework and the details of RSMS are given in Sections III and IV, respectively. Reliability and security are analyzed in Section V. We present performance analysis in Section VI. Finally, the conclusion and future work of this paper are presented in Section VII.

## II. RELATED WORK

This section introduces the related works from aspects of Metaverse service security and Metaverse service quality of service (QoS) and quality of experience (QoE).

### A. Metaverse Service Security

Security threats in the Metaverse service primarily revolve around various dimensions, such as authentication, access management, data handling, privacy concerns, network vulnerabilities, economic risks, physical implications, and governance issues [14]. Ensuring the integrity and authentication within the Metaverse stands as a paramount challenge [15]. Unauthorized identity theft and impersonation of users/avatars are concerning issues in the Metaverse, and complications arise when attempting to establish authentication compatibility across various virtual worlds. Researchers have undertaken investigations in the field of authentication and access control to address these challenges. Sethuraman *et al.* [16] proposed a passwordless authentication method with SDKs for seamless authentication across systems, including VR/XR glasses, enhancing Metaverse service accessibility. Yang *et al.* [17] presented a two-factor authentication framework that relied on biometric-based authentication combined with a chameleon signature approach. Tuan *et al.* [18] introduced BlockBee, a Blockchain-based architecture for decentralized authentication and access control for smart devices in SDN-supported networks within the Metaverse context. Sethuraman *et al.* [19] introduced MetaKey, a novel framework for authentication and identity management designed to enhance the security of digital assets and digital identity. Lai *et al.* [20] presented a blockchain-based multi-factor group authentication scheme tailored for Metaverse scenarios. Cheng *et al.* [21] formulated a comprehensive research agenda for zero-trust user authentication in social VR, an early prototype of the Metaverse.

However, although the above authentication schemes are designed to ensure the security of Metaverse service, they either do not support cross-domain authentication and group authentication at the same time, or do not meet all security attributes (see Section III.D.2), especially forward and backward security, which are required when authenticating RNs in Metaverse service provision.

### B. Metaverse Service QoS&QoE

There are still some works on the Metaverse QoS and QoE. The works [22][23] investigated data availability in the Metaverse concerning data synchronization and QoS, respectively. For precise DT synchronization with the physical counterpart, Han *et al.* [23] proposed a hierarchical game for dynamic DT synchronization in the Metaverse. In this approach, end devices collectively collected status information from physical objects, and virtual service providers (VSPs) determined suitable synchronization intensities. Utilizing covert communication methods, Du *et al.* [22] suggested an optimal targeted advertising strategy for VSPs to maximize its payoff in delivering high-quality access services to end-users while achieving nearly flawless detection errors for potential attackers. Chua *et al.* [24] introduced a Metaverse-compatible Unified User and Task (UUT)-centered AI-based Mobile Edge Computing (MEC) paradigm. This concept serves as a foundation for future AI control algorithms aimed at creating a more user and task-centric MEC. Ishimaru *et al.* [25] presented a point cloud streaming method for real-time 3D reconstruction in VR space, enabling streaming of real-space objects like humans and animals while optimizing the user's QoE within resource constraints (computational and network resources). Taking a consumer-centric approach, Du *et al.* [26] explored the interplay between Metaverse system design and consumer behaviors. They redefined QoE and introduced a framework that integrated both Metaverse service providers (MSPs) and consumer perspectives.

Different from the above works, we study the reliability of Metaverse service. Our work can complement these existing works for better Metaverse service provision.

## III. PRELIMINARIES

This section first introduces background knowledge. Then the system framework and Metaverse service requirements are given.

### A. Background Knowledge

**Definition 1.** (Elliptic Curve [27]): An elliptic curve $E$ over a prime finite field $F_p$ is defined by the equation $y^2 = x^3 + ax + b \bmod p$, where $a,b \in F_p$ and $4a^3 + 27b^2 \neq 0$. All the points on $E$ with the point at infinity $O$ form the corresponding elliptic curve group $G = \{(x,y): x,y \in F_p, E(x,y)=0\} \cup \{O\}$ regarding scalar multiplication and point addition.

- Elliptic Curve Discrete Logarithm Problem (ECDLP): Given $P, Q \in G$, it is computationally hard for any polynomial-time bounded algorithm to determine $x$ such that $x \in Z_q^*$ and $Q = xP$.
- Elliptic Curve Diffie-Hellman Problem (ECDHP): Given $P \in G, x, y \in Z_q^*$, it is computationally hard for any polynomial-time bounded algorithm to determine $xyP$ such that $X = xP, Y = yP$.

**Definition 2.** (Bloom Filter [28]): The Bloom Filter (BF) is a probabilistic data structure that not only provides space-efficient storage of a set but also can efficiently test whether an element is a member of the set. The probabilistic property of BF may lead to false positive matches, but not false negatives. The more elements are in the BF, the higher the chance of getting a false positive match insertion. To reduce its false positive rate, this paper follows the approach of [29], i.e., a BF with $1.44\varepsilon N$ bits for a set with size $N$ has a false positive rate (FPR) of $2^{-\varepsilon}$.

The key components of the BF algorithm are as follows:
- $Init(N, \varepsilon)$: On input, a set size $N$, the initialization algorithm initiates the BF of bit length $1.44\varepsilon N$.
- $Insert(m)$: Element insertion algorithm takes an element $m$ as input, and inserts $m$ into BF.
- $Check(m)$: Element check algorithm returns 1 if an element $m$ is in BF, and 0 otherwise.
- $Pos(m)$: Position update algorithm computes positions to be changed for element $m$ in BF.

**Definition 3.** (Shamir Secret Sharing [30]): A secret sharing scheme over $Z_q$ is composed of two algorithms, namely, $G$ and $C$. $G$ is a probabilistic algorithm that generates $t$-out-of-$n$ shares of $s$. $G$ is invoked as $G(n, yt, s) \xrightarrow{R} (s_1', s_1', ..., s_t')$ where $n$ is the number of shares, $t$ is the threshold such that $(0 < t \leq n)$, $k$ is the secret, and $s_i$ is the share for node $i$. $C$ is a deterministic algorithm $s \longleftarrow C(s_1', s_1', ..., s_t')$. It is invoked to recover $s$ using the Lagrange's interpolation formula. And for every $t$ set of shares of $s$, $C(s_1', s_2', ..., s_t') = s$.

### B. System Framework

Fig.2 describes the system framework of Metaverse service considered in this paper, which is composed of a MSRP, several RPMs, a dense set of RNs, and Users. The detailed descriptions of these entities are as follows.

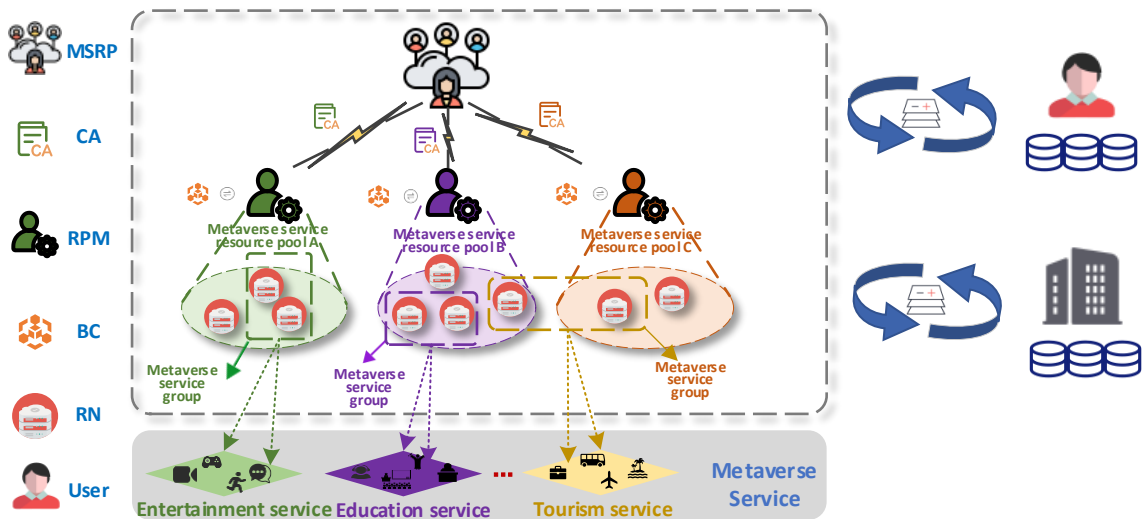

Fig.2. System Framework

**Metaverse Service Resource Provider (MSRP):** The MSRP is fully trusted. It owns a set of public and private key pairs, and assumes the responsibility of generating public security parameters. It is also responsible for registering RPMs, RNs, and Users, allocating RNs to Users for handling their Metaverse service requests.

**Resource Pool Manager (RPM):** Each RPM is responsible for managing its Metaverse service resource pool, including the initial authentication and re-authentication of RNs joining its Metaverse service resource pool. Besides, it is also responsible for forwarding information such as computing or storage resources of RNs within its Metaverse service resource pool to the MSRP. Moreover, all RPMs maintain a shared blockchain to store information related to RNs in its Metaverse service resource pool.

**Resource Node (RN):** Each RN has computing and/or storage resources. When joining a Metaverse service resource pool, it first must register through MSRP to obtain its public- private key pair. Then it needs to be authenticated by the RPM, which is in charge of the Metaverse service resource pool it chooses. When the RN is idle in the current Metaverse service resource pool, it can join another Metaverse service resource pool, which desperately needs resources, for more profits. Before joining, it also needs to be authenticated by the corresponding RPM of the new Metaverse service resource pool. When providing Metaverse service resources to the User, it needs to be authenticated by the User and communicates with other RNs within the same Metaverse service group.

**User:** The User requests the Metaverse service, which are provided by allocating RNs, to deploy Metaverse service. Before that, the User must register through the MSRP and then sends a request to the MSRP. After that, to ensure reliable and secure Metaverse service that the User deploys on RNs, the group authentication protocol must be done among the User and the allocated RNs.

### C. Security Model

In the security model, there are two types of adversaries: $\mathcal{A}_\mathrm{I}$ and $\mathcal{A}_\mathrm{II}$. They attack the unforgeability and confidentiality of RSMS, respectively.

**Definition 4: Unforgeability.** RSMS for RN joining a Metaverse service resource pool meets unforgeability if the probability that the adversary $\mathcal{A}_\mathrm{I}$ can solve the ECDLP is negligible in any polynomial time. In RSMS, the unforgeability is defined as existential unforgeability under chosen message attack (EUF-CMA).

**Definition 5: Confidentiality.** RSMS for User accessing Metaverse service resources meets confidentiality if the probability that the adversary $\mathcal{A}_\mathrm{II}$ could solve the ECDLP is negligible in any polynomial time. In RSMS, confidentiality is defined as indistinguishability under the adaptive-chosen-ciphertext attack (IND-CCA2).

### D. Requirements of Metaverse Service

#### D.1) Reliability

**Continuity of Metaverse service.** This requirement essentially stresses the importance of maintaining Metaverse service. To be specific, due to some reasons, there is a possibility of an RN being offline or malfunctioning during Metaverse service provision, then one or more new RNs will join the corresponding Metaverse service group to provide Metaverse service to the User. In this process, the new RNs must be authenticated and the group session key between User and RNs must be updated.

#### D.2) Security

Seven security attributes are detailed in the following.

**Confidentiality.** Messages are transmitted via public channels when RNs provide Metaverse service resources. Confidentiality requires that it is computationally infeasible for adversaries to attempt to decrypt these transmitted messages.

**Integrity.** Entities can verify the integrity of the transmitted messages and discover any changed message.

**Non-repudiation.** Upon message reception, the recipient is capable of ascertaining the identity of the sender, thereby preventing the sender from repudiating their role in transmitting said message.

**Anonymity.** The identity information of entities, including RPMs, RNs and Users, must be protected when RNs provide Metaverse service resources.

**Traceability.** When an RN joins a Metaverse service resource pool, the information related to it must be recorded for trace and audit.

**Unlinkability.** RN's pseudonyms stored in the blockchain must be unlinkability each other in order to prevent RN's privacy leakaging caused by link attacks.

**Forward security and backward security.** A new RN joining the Metaverse service group must satisfy forward security and backward security. To be specific, in this paper, forward security means that the suspicious RN can't know any new information about the deployed Metaverse service after it quits. And backward security means that the new RN must be authenticated before it joins the Metaverse service group.

## IV. RELIABLE AND SECURE METAVERSE SERVICE PROVISION MECHANISM

In this section, we detail RSMS, which contains two protocols: one is used for RN joining a Metaverse service resource pool and the other is for settting up and maintaining a stable and secure group of RNs for service provision. Notations to be used in the following are described in TABLE I.

### A. RN Joining a Metaverse Service Resource Pool

This section details the process of RN joining a Metaverse service resource pool. The process mainly includes two phases, initialzation phase and mutual authentication phase. The workflow is shown in Fig.3.

#### A.1) Initialization Phase

Initialization phase aims to initialize the whole system and prepare for authentication, including RPM Setup, RN Setup, and RN Setup. The details are as follows.

**MSRP Setup:** Given the security parameter $\kappa \in Z^+$ as input, *MSRP* generates the public parameters.

1) First, *MSRP* chooses an Elliptic curve $E: y^3 = x^3 + ax + b \bmod p$ and obtains the group $G$ of prime order $q$ with a generator $P$.

2) Secondly, it chooses four secure cryptographic hash functions, $H_1: G \to \{0,1\}^*, H_2: \{0,1\}^* \to Z_q^*, H_3: G \times \{0,1\}^* \to Z_q^*, H_4: Z_q^* \to Z_q^*, H_5: G \to G, H_6: \{0,1\}^* \times Z_q^* \to \{0,1\}^*$.

TABLE I  DEFINITION OF SYMBOLS

| Notation | Description |
|---|---|
| $BF^X$ | Bloom filter in a Metaverse service resource pool $X$ |
| $RPM^X$ | RPM in a Metaverse service resource pool $X$ |
| $RN_i^X$ | i-th RN in a Metaverse service resource pool $X$ |
| $RID_{RPM}^X$ | Real identity of $RPM^X$ |
| $PID_{RPM}^X$ | Pseudonym of $RPM^X$ |
| $RID_{RN_i}^X$ | Real identity of $RN_i^X$ |
| $PID_{RN_{i,k}}^X$ | k-th pseudonym of $RN_i^X$ |
| $msk$ | Master secret key of MSRP |
| $PK_{pub}$ | Master public key of MSRP |
| $sk_{RPM}^X$ | Secret key of $RPM^X$ |
| $PK_{RPM}^X$ | Public key of $RPM^X$ |
| $sk_{RN_i}^X$ | Secret key of $RN_i^X$ |
| $PK_{RN_i}^X$ | Public key of $RN_i^X$ |
| $Enc_{PK}(m)$ | Encrypt m by PK using ECC |
| $SRN$ | Set of suspicious $RNs$ |
| $s_{serv}$ | Group Session key |

3) Then, it randomly chooses $msk \in Z_q^*$, and calculates $PK_{pub} = msk \cdot P$.

4) Additionally, it uses $Init(N, \varepsilon)$ to initialize a BF, $BF^X$.

5) Finally, *MSRP* publishes the public parameters $params = \{q, p, G, P, PK_{pub}, H_1, H_2, H_3, H_4, H_5, H_6\}$ and keeps the master secret key $msk$ secretly.

**RPM Setup:** The setup of $RPM^X$ means that it obtains the legitimate public-private key pair $(sk_{RPM}^X, PK_{RPM}^X)$, which are used in mutual authentication phase.

1) $RPM^X \longrightarrow MSRP: \{Enc_{PK_{pub}}(RID_{RPM}^X)\}$. $RPM^X$ encrypts $RID_{RPM}^X$ using the public key $PK_{pub}$, and then sends it to *MSRP*.

2) $MSRP \longrightarrow RPM^X: \{(sk_{RPM}^X, PK_{RPM}^X), PID_{RPM}^X\}$. After receiving the encrypted message, *MSRP* decrypts it by using $msk$. Then *MSRP* chooses a number $d \in Z_q^*$ randomly and computes the public key $PK_{RPM}^X = d \cdot P$ and the hash value $PID_{RPM}^X = H_1(PK_{pub}) \oplus RID_{RPM}^X$. Besides, it computes the secret key $sk_{RPM}^X = d + msk \cdot H_2(PID_{RPM}^X)$. Then *MSRP* sends $(sk_{RPM}^X, PK_{RPM}^X)$ and $PID_{RPM}^X$ to $RPM^X$ via a secure channel.

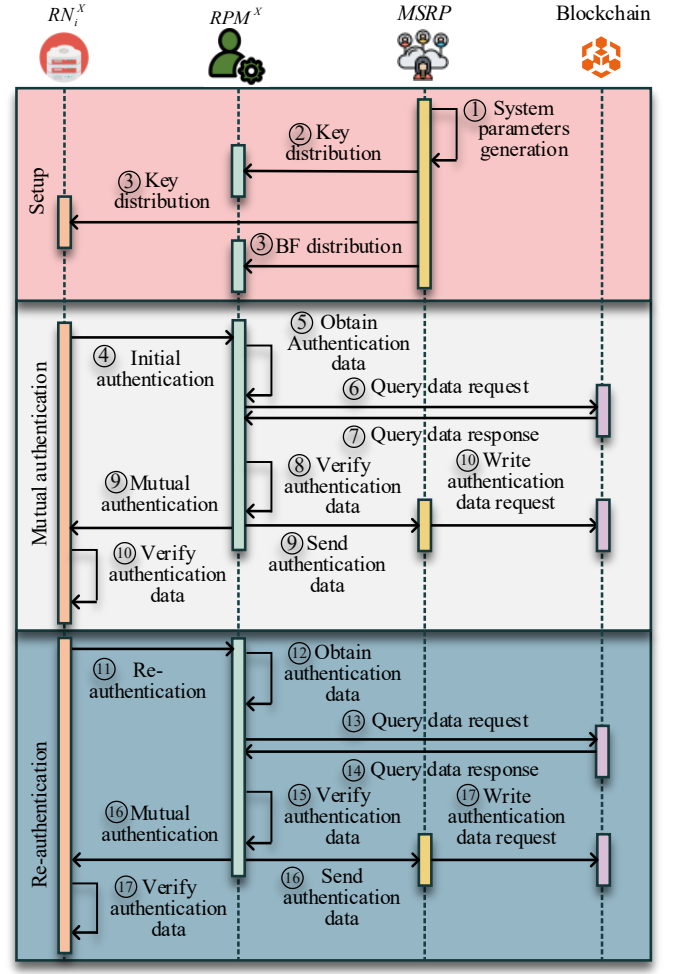

Fig.3. Workflow of RN Joining a Metaverse Service Resource Pool

**RN Setup:** The setup of $RN_i^X$ is similar to that of $RPM^X$. Besides, $RN_i^X$ will obtain extra data used in mutual authentication phase.

1) $RN_i^X \longrightarrow MSRP: \{Enc_{PK_{pub}}(RID_{RN_i}^X)\}$. $RN_i^X$ encrypts $RID_{RN_i}^X$ using the public key $PK_{pub}$, and then sends it to *MSRP*.

2) $MSRP \longrightarrow RN_i^X: \{sk_{RN_i}^X, PK_{RN_i}^X, a_{RN_{i,k}}^X, PID_{RN_{i,k}}^X\}$. After receiving the encrypted message, *MSRP* decrypts it using $msk$ and obtain $RID_{RN_i}^X$. Then *MSRP* chooses a random number $sk_{RN_i}^X \in Z_q^*$ and computes the public key $PK_{RN_i}^X = sk_{RN_i}^X \cdot P$. Additionally, *MSRP* also chooses a random number $a_{RN_{i,k}}^X \in Z_q^*$ randomly and computes the hash value $PID_{RN_i}^X = H_1(PK_{pub}) \oplus RID_{RN_i}^X$. It also uses $Insert(m)$ to insert $a_{RN_{i,k}}^X$ into the $BF^X$. Finally, *MSRP* sends $(sk_{RN_i}^X, PK_{RN_i}^X)$, $a_{RN_{i,k}}^X$ and $PID_{RN_{i,k}}^X$ to $RN_i^X$ via a secure channel.

Note that during the setup of $RN_i^X$, MSRP will send $BF^X$ to the RPM, which is in charge of the Metaverse service resource pool *X*, via a secure channel when enough RNs have been registered, or a certain interval of time has passed. Then *MSRP* will initialize a new

$BF^X$.

*A.2) Mutual Authentication Phase*

In mutual authentication phase, $RN_i^X$ needs to complete the mutual authentication with $RPM^X$, including **Initial Authentication** and **Re-authentication**. And the **Re-authentication** is used to realize the authentication of RNs across Metaverse service resource pools.

*a) Initial Authentication*

**Initial Authentication:** It is required when $RN_i^X$ firstly connects to Metaverse. The process is described as follows.

1) $RN_i^X \longrightarrow RPM^X : \{c, PID_{RN_{i,k}}^X\}$. $RN_i^X$ chooses a random number $u \in Z_q^*$ and computes $U = u \cdot P$. Then $RN_i^X$ computes $c_1 = u(PK_{RPM}^X + H_2(PID_{RPM}^X) \cdot PK_{pub})$. After that, $RN_i^X$ computes $c_2 = H_3(U, PID_{RN_{i,k}}^X) \oplus a_{RN_{i,k}}^X$. Finally, $RN_i^X$ sends $c = (c_1, c_2)$ and $PID_{RN_{i,k}}^X$ to $RPM^X$.

2) $RPM^X \longrightarrow MSRP : \{Enc_{PK_{pub}}(a_{RN_{i,k}}^X{}', PID_{RN_{i,k}}^X)\}$. After receiving the message, $RPM^X$ computes $U' = c_1' \cdot sk_{RPM}^{X^{-1}}$. Then, $RPM^X$ computes $a_{RN_{i,k}}^X{}' = H_3(U', PID_{RN_{i,k}}^X) \oplus c_2'$. After that, $RPM^X$ firstly searches if $H_4(a_{RN_{i,k}}^X{}')$ is in blockchain. If not, it uses $Check(m)$ to check if $a_{RN_{i,k}}^X{}'$ is in $BF^X$. If true, $RPM^X$ encrypts $a_{RN_{i,k}}^X{}'$, $PID_{RN_{i,k}}^X$ using $PK_{pub}$ and sends it to $MSRP$. Otherwise, it rejects the $RN_i^X$.

3) $RPM^X \longrightarrow RN_i^X : \{A_{RN_{i,k}}^X\}$. After finishing authentication of $RN_i^X$, $RPM^X$ computes $A_{RN_{i,k}}^X = a_{RN_{i,k}}^X \cdot P$ and sends it to $RN_i^X$.

4) After receiving the message, $RN_i^X$ verifies it through the equation $a_{RN_{i,k}}^X \cdot P = A_{RN_{i,k}}^X{}'$. If true, $RN_i^X$ completes the authentication of $RPM^X$.

Then $RN_i^X$ completes **Initial Authentication** and can join the Metaverse service resource pool $X$. Note that when $MSRP$ receives the message from $RPM^X$ in step 2 of **Initial Authentication**, it will verify and storage it in the blockchain. Firstly, $MSRP$ will verify if $a_{RN_{i,k}}^X$ is valid. If valid, $MSRP$ updates $a_{RN_{i,k}}^X$ and $PID_{RN_{i,k}}^X$ for the corresponding $RN_i^X$ by computing $a_{RN_{i,k+1}}^X = H_4(a_{RN_{i,k}}^X, sk_{RN_i}^X)$ and $PID_{RN_{i,k+1}}^X = H_6(PID_{RN_{i,k}}^X, a_{RN_{i,k+1}}^X)$. And then it computes $H_4(a_{RN_{i,k}}^X)$ and storages it with $PID_{RN_{i,k}}^X$ in the blockchain.

Due to $PK_{RPM}^X = d \cdot P$, $sk_{RPM}^X = d + msk \cdot H_2(PID_{RPM}^X)$, and $PK_{pub} = msk \cdot P$, $RPM^X$ can obtain the correct $U$ by using the below formula.

$$U' = c_1' \cdot sk_{RPM}^{X^{-1}}$$
$$= u(PK_{RPM}^X + H_2(PID_{RPM}^X) \cdot PK_{pub}) \cdot sk_{RPM}^{X^{-1}}$$
$$= u(d \cdot sk_{RPM}^X + sk_{RPM}^X \cdot P - d \cdot sk_{RPM}^X) \cdot sk_{RPM}^{X^{-1}}$$
$$= u \cdot P$$
$$= U$$

**Remark 1.** $sk_{RPM}^{X^{-1}}$ can be precomputed and the verification of $a_{RN_{i,k}}^X$ is done by using BF, which is more efficient. Therefore, during **Initial Authentication**, the computational overhead of $RPM^X$ is small. And step 2 and step 3 can also be parallel in order to save time.

*b) Re-authentication*

**Re-authentication:** $RN_i^X$ sometimes may join other Metaverse service resource pools due to various reaseons. Then it needs to communicate with another RPM, denoted as $RPM^Y$. And no matter which situation happens, $RN_i^X$ must be re-authenticated. The details are as follows.

1) $RN_i^X \longrightarrow RPM^Y : \{c, PID_{RN_{i,k+1}}^X\}$. $RN_i^X$ computes $a_{RN_{i,k+1}}^X = H_4(a_{RN_{i,k}}^X, sk_{RN_i}^X)$, and $PID_{RN_{i,k+1}}^X = H_6(PID_{RN_{i,k}}^X, a_{RN_{i,k+1}}^X)$. $RN_i^X$ chooses a random number $w \in Z_q^*$ and computes $W = w \cdot P$. Then $RN_i^X$ computes $c_1 = w(PK_{RPM}^Y + H_2(PID_{RPM}^Y) \cdot PK_{pub})$. After that, $RN_i^X$ computes $c_2 = H_3(W, PID_{RN_{i,k}}^X) \oplus (H_4(a_{RN_{i,k}}^X) \| a_{RN_{i,k+1}}^X)$. Finally, $RN_i^X$ sends $c = (c_1, c_2)$, $PID_{RN_{i,k}}^X$ and $PID_{RN_{i,k+1}}^X$ to $SP^Y$.

2) $RPM^Y \longrightarrow MSRP : \{Enc_{PK_{pub}}(a_{RN_{i,k+1}}^X{}'), PID_{RN_{i,k+1}}^X\}$. After receiving the message, $RPM^Y$ computes $W' = c_1' \cdot sk_{RPM}^{Y^{-1}}$. Then, $RPM^Y$ computes $H_4(a_{RN_{i,k}}^X{}') \| a_{RN_{i,k+1}}^X{}' = H_3(U', PID_{RN_{i,k}}^X) \oplus c_2'$. After that, $RPM^Y$ firstly searches if $PID_{RN_{i,k+1}}^X$ is not in blockchain and $H_4(a_{RN_{i,k}}^X{}')$ is in blockchain. If true, $RPM^Y$ encrypts $a_{RN_{i,k+1}}^X{}'$ using $PK_{pub}$ and sends it with $PID_{RN_{i,k+1}}^X$ to $MSRP$. Otherwise, it rejects the access request.

3) $RPM^Y \longrightarrow RN_i^X : \{A_{RN_{i,k+1}}^X\}$ After finishing the authentication of $RN_i^X$, $RPM^Y$ computes $A_{RN_{i,k+1}}^X = a_{RN_{i,k+1}}^X \cdot P$ and sends it to $RN_i^X$.

4) After receiving the message, $RN_i^X$ verifies it through the equation $a_{RN_{i,k+1}}^X \cdot P = A_{RN_{i,k+1}}^X{}'$. If true, $RN_i^X$ completes the authentication of $RPM^Y$.

Then $RN_i^X$ completes the **Re-authentication** and can join the Metaverse service resource pool $Y$. Note that when $MSRP$ receives the message from $RN_i^X$ in step 2 of **Re-authentication**, it will do things, which are the same as the things in **Initial Authentication**.

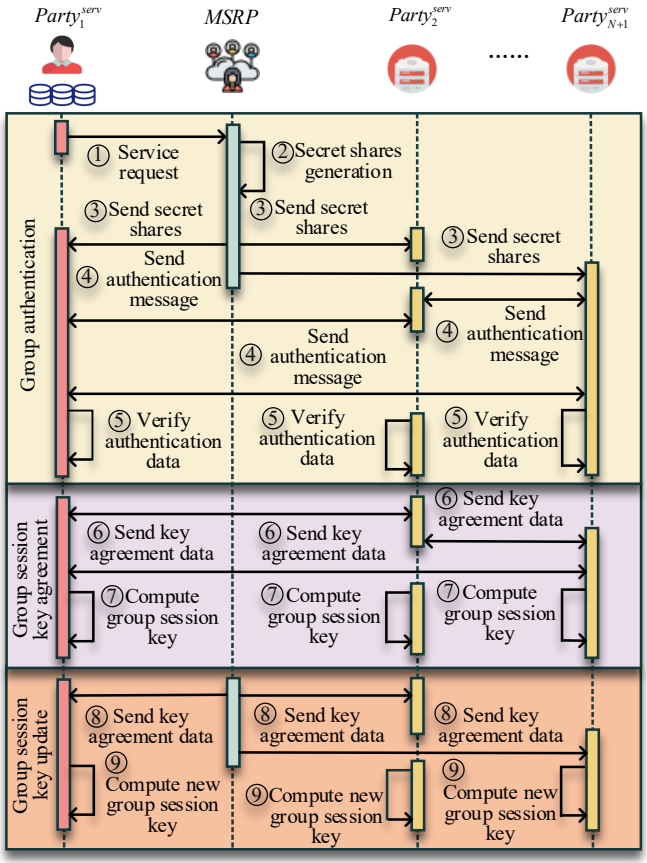

Fig.4. Workflow of RN Joining a Metaverse Service Group

## B. RN Joining a Metaverse Service Group

After RNs accomplish the joining Metaverse resource pool authentication, they can provide Metaverse service resources for Users. Before that, the User needs to establish connection with RNs assigned to it. This process mainly contains two phases: user register phase and group authentication phase, which are showed in Fig.4.

*B.1) User Register Phase*

In this phase, the User needs to communicate with MSRP in order to request Metaverse service resources. The details are as followed.

**User Register**. The User obtains the legitimate public-private key pair $(sk_{User}, PK_{User})$ and $PID_{User}$, which are used in the group authentication phase. The steps are similar to RN setup. In addition, MSRP also maintains User's $(sk_{User}, PK_{User})$ and $PID_{User}$.

*B.2) Group Authentication Phase*

When the User wants to request Metaverse service resources, it sends the request to MSRP. Then MSRP will allocate several RNs to it. And the User needs to complete the group authentication protocol with these RNs and obtain the group session key. Note that messages sent to RNs by MSRP need be forwarded by the RPM. The group authentication protocol includes three subprotocols, **Group Membership Authentication, Group Session Key Agreement, Group Session Key Update.**

*a) Group membership authentication*

**Group Membership Authentication:** Before negotiating the group session key, **Group Membership Authentication** among User and RNs must be done. In order to simplify the denotation, we use $Party_1^{serv}$ to represent the User and $Party_2^{serv}, Party_3^{serv}, ...$ to represent RNs in the rest of the paper. The detailed description are as followed.

1) $Party_1^{serv} \longrightarrow MSRP : \{serv\ request\}$. $Party_1^{serv}$ sends the computing resources request to $MSRP$.

2) $MSRP \longrightarrow Party_i^{serv} (1 \leq i \leq N+1) : \{Enc_{PK_{Party_i^{serv}}}(f(x_i)), List_{PID_i^{serv}}\}$. $MSRP$ allocates $N$ RNs according to the User's request. Then it chooses $r \in Z_q^*$ and a polynomial $f(x) = s_{serv} + a_1 x + ... + a_N x^N$ with degree $N$ and computes $s_{serv} = r + msk H_2(PID_i^{serv} \| ...)(1 \leq i \leq N+1)$ and $Q = s_{serv} \cdot P$. After that, $MSRP$ computes $x_i = H_2(PID_i^{serv})$ and $f(x_i)$ for $Party_i^{serv} (1 \leq i \leq N+1)$. Finally, $MSRP$ encrypts $f(x_i)$ with $PK_{Party_i}$ and sends the result with $List_{PID_i^{serv}}$, which contians the $PID$ of every $Party$, to $Party_i^{serv}$ for $i=1,...,N+1$.

3) $Party_i^{serv} \longrightarrow Party_j^{serv} (j \neq i) : \{f(x_i) \cdot P\}$. $Party_i^{serv}$ decrypts the received message, obtains $f(x_i)$, and computes the $f(x_i) \cdot P$. Then $Party_i^{serv}$ sends $f(x_i) \cdot P$ to $Party_j^{serv} (j \neq i)$.

4) $Party_i^{serv} \longrightarrow Party_j^{serv} (j \neq i) : \{C_i\}$. $Party_i^{serv}$ computes
$$C_i = \left( \prod_{r=1, r\neq i}^{N+1} \frac{-x_r}{x_i - x_r} \right) f(x_i) \cdot P.$$

5) $Party_i^{serv}$ verify $\sum_{i=1}^{N+1} C_i = Q$.

If true, Group Membership Authentication is completed.

*b) Group session key agreement*

**Group Session Key Agreement**: Once the User and RNs complete **Group Membership Authentication**, they will conduct the **Group Session Key Agreement**. The details are described as follow.

1) $Party_i^{serv} \longrightarrow Party_j^{serv} (j \neq i) : \{f(x_i)f(x_j) \cdot P \oplus f(x_i)\}$. $Party_i^{serv}$ computes $f(x_j)f(x_i) \cdot P (j \neq i)$. Then it sends the result to $Party_j^X (j \neq i)$.

2) $Party_j^{serv}$ computes $f(x_j)f(x_i) \cdot P (j \neq i)$ and $f(x_i)f(x_j) \cdot P \oplus f(x_i) \oplus f(x_j)f(x_i) \cdot P$. Then it obtains $f(x_i)$. After that, $Party_j^{serv}$ computes $s_{serv}' = \sum_{i=1}^{m} f(x_j) \prod_{r=1, r \neq j}^{m} \frac{-x_r}{x_j - x_r}$.

Finally, $Party_j^{serv}$ verifies whether $H_4(s_{serv}') = H_4(s_{serv})$. If true, it obtains the group session key $s_{serv}$.

Then the User can use the allocated RNs to deploy its Metaverse service.

*c) Group session key update*

**Group Session Key Update:** Note that RNs will join and exit Metaverse service group stochasticly. For example, during the RN providing Metaverse service resources for Users, it is detected for malicious behaviors or it suffers from software aging. Then it leaves the group actively or passively. After that, the RPM will allocate new RNs to join the Metaverse service group. In order to guarantee forward and backward security, the User and Metaverse service group must support **Group Session Key Update**. And the details are described as follows.

1) $MSRP \longrightarrow Party_i^{serv}(i \notin SRN):\{Enc_{PK_i^{serv}}(s_{serv}^{new})\}$. If $MSRP$ finds $Party_k^{serv}(1 < k \le N+1, k \in M)$ are suspicious. It will compute $s_{serv}^{new} = s_{new} + msk H_5(Q, PK_{pub})$. Then $MSRP$ encrypts $s_{serv}^{new}$ and sends the result to $Party_i^{serv}(i \notin M)$.

2) $Party_i^{serv}(i \notin M)$ will decrypt the message and verify if $s_{serv}^{new'} \cdot P = Q + H_5(Q, PK_{pub}) \cdot PK_{pub}$. If true, it obtains the new group session key.

## V. RELIABILITY AND SECURITY ANALYSIS

This section discusses the capability of RSMS in assuring reliability and security of Metaverse service. In the following, we use a general $UID$ to represent one role in $\{RPM, RN, User\}$ and a random oracle. $\mathcal{S}$ simulates oracles queried by $\mathcal{A}_I$ or $\mathcal{A}_{II}$.

**Game Unforgeability.** The game is an interaction between the simulator $\mathcal{S}$ and the adversary $\mathcal{A}_I$ under IND-CCA2. The specific definition of this game is as follows:

- *Hash* query returns the value $Hash(m)$ if $m$ exists in the hash list. Otherwise, a random value will be generated, added to the hash list, and returned. Let the list $L_{H_i}$ store the answers of random oracle $H_i(i=1,2,3,4,5,6)$. For a hash oracle query $H_i(m)$, if a record $(m, h_i)$ exists in $L_{H_i}$, return $h_i$. Otherwise, a random value $h_i$ will be chosen, added $(m, h_i)$ to $L_{H_i}$, and returned.

- *Send*(M, UID) query simulates the active attacks, in which $\mathcal{A}_I$ can modify the transmitted messages via the public channel. The random oracle $UID$ replies to the query if $\mathcal{A}_I$ sends the modified message $M$ to $UID$. For a *Send*($Mes_1$, $RPM$), answer it as follows. Decrypt the message by computing $U' = c_1' \cdot sk_{RPM}^{X^{-1}}$ and $a_{RN_{i,k}}^X{}' = H_3(U', PID_{RN_{i,k}}^X) \oplus c_2'$. Then, search if $H_4(a_{RN_{i,k}}^X{}')$ is in blockchain or if $a_{RN_{i,k}}^X{}'$ is not in $BF^X$. If true, abort the session. Otherwise, the RN is authenticated. Finally, return $A_{RN_{i,k}}^X = a_{RN_{i,k}}^X \cdot P$. For a *Send*($Mes_2$, $RN$) query, answer it as follows. Check if $a_{RN_{i,k}}^X \cdot P = A_{RN_{i,k}}^X{}'$. If it is not valid, abort the session. Otherwise, return message encrypted by $PK_{pub}$.

- *Execute*(RPM, RN) query simulates the passive attacks, in which $\mathcal{A}_I$ can eavesdrop on the channel and learn the transmitted messages between RPM and RN. For an *Execute*(RPM, RN) query, it proceeds with the simulation of the above defined *Send* queries successively, and returns ($Mes_1$, $Mes_2$) which has been transmitted over the public channel.

- *IDReveal*(UID) query allows $\mathcal{A}_I$ to know the PID of RPM or RN. For a *IDReveal*(RPM) or a *IDReveal*(RN) query, answer it by execute *H*-query. Then, return the output of *H*-query.

Forgery: After the above queries are over, adversary $\mathcal{A}_I$ returns a forged signature pair $(m, \sigma)$. If the signature pair is valid, then $\mathcal{A}_I$ wins **Game Unforgeability**; otherwise, $\mathcal{A}_I$ fails. Note that in the forgery process, $\mathcal{A}_I$ cannot query for the signature key.

**Game Confidentiality:** The game is an interaction between the simulator $\mathcal{S}$ and the adversary $\mathcal{A}_{II}$ under IND-CCA2. The specific definition of this game is as follows:

- *Hash* query perform the same operations as in **Game Unforgeability**.

- *Send*(M, UID) query simulates the active attacks, in which $\mathcal{A}_{II}$ can modify the transmitted messages between party and party. For a $Send(Mes_1, Party_i)$ query, answer it as follows. First, compute $C_i = \left(\prod_{r=1, r \ne i}^{N+1} \frac{-x_r}{x_i - x_r}\right) f(x_i) \cdot P$ and verify $\sum_{i=1}^{N+1} C_i = Q$. If not, abort the session. Otherwise, compute $f(x_i)f(x_j) \cdot P(j \ne i)$ and return it.

- *Execute*($Party_i$, $Party_j$) query simulates the passive attacks, in which $\mathcal{A}_{II}$ can eavesdrop on the channel and learn the transmitted messages among the Metaverse service group. For an *Execute*($Party_i$, $Party_j$) query, it proceeds with the simulation of the above defined *Send* queries successively, and returns $Mes_1$ which has been transmitted over the public channel.

- *GSKReveal*(UID) query permits $\mathcal{A}_{II}$ to learn the the session key generated by the random oracle $UID$. For a *GSKReveal*(UID) query, answer with the session key $s_{serv}$ if the target instance has actually generated a session key, and both $UID$ and its partner are not asked by a *Test* query. Otherwise, *null* is returned.

- *Corrupt*(UID) query allows $\mathcal{A}_{II}$ to learn the the long-term private key of the entity $UID$. For a *Corrupt*($Party_i$) query, answer the query with the long-term private key of the entity $Party_i$.

- *Test*(UID) query returns a session key or a random value to $\mathcal{A}_{II}$ by the random oracle $UID$. Note that this query cannot be issued when *GSKReveal* and *Corrupt* querries have been executed. For a *Test*(UID) query, it first gets $s_{serv}$ from *GSKReveal*(UID) query, and then flips a coin $b$. If $b=1$, the session key $s_{serv}$ is returned. Otherwise, a random value chosen from $\{0,1\}^l$ is returned.

Challenge: After the above queries are over, adversary $\mathcal{A}_{II}$ sends two plaintexts $\{m_0, m_1\}$ to $\mathcal{S}$. The $\mathcal{S}$ randomly chooses a bit $b \in \{0,1\}$, calculates the encryption message $\delta$ corresponding to $m_d$, and final returns the $\delta$ to $\mathcal{A}_{II}$. $\mathcal{A}_{II}$ ask $\mathcal{S}$ some queries. However, the $\mathcal{A}_{II}$ cannot query for the decryption key of $\delta$ and cannot query for the decryption result of $\delta$. Finally, the $\mathcal{A}_{II}$ outputs a bit $b' \in \{0,1\}$. If $b' = b$, $\mathcal{A}_{II}$ wins the **Game Confidentiality**, otherwise, $\mathcal{A}_{II}$ fails.

*A. Formal Analysis*

We first give the difference lemma (*Lemma 1*), developed in [32]. Then *Theorem 1* and *Theorem 2* are present.

*Lemma 1.* (Difference Lemma): Let $E_1, E_2$ and $E_3$ denote the events which follow a general probability distribution. If $E_1 \wedge \neg E_3 \Leftrightarrow E_2 \wedge \neg E_3$, we have $|\Pr[E_1] - \Pr[E_2]| \leq \Pr[E_3]$.

*Theorem 1.* If, for any polynomial time adversary $\mathcal{A}_I$, $Adv_{\mathcal{A}_I}^{RSMS} < \frac{O(q_h)^2}{2^l} + O(q_h T) + 2^{-\varepsilon}$, where $q_h$ is a maximum of *Hash* queries, then RN joining a Metaverse service resource pool can be considered as satisfying unforgeability. Note that $\kappa$ is the given security parameter and $T_1$ is the time of solving the ECDLP.

*Proof.* We show the proof of **Theorem 1** in a sequence of games following the approach described in [32]. The success probability of $\mathcal{A}_I$ only increases by a negligible amount when moving between the games, as a consequence of *Lemma 1*.

**Game 0**: This game follows the original process, which is called **Game Unforeability**. Meanwhile, all queries are answered honestly according to the specification of RN joining a Metaverse service resource pool. Thus, we have that
$$Adv_{\mathcal{A}_I}^{RSMS}(k) = Adv_0$$

**Game 1**: This game proceeds like **Game 0**, but avoids some collisions in the output of hash functions, including $PID_{SP}^X, PID_{RN_i}^X, a_{RN_{i,k}}^X$, and the selection of random values $u$ by a participant *UID*. And we assume that they are statistically indistinguishable from random strings under random oracle model. According to **Lemma 1**, it holds that
$$|Adv_0 - Adv_1| \leq \frac{O(q_h)^2}{2^l}$$

**Game 2**: In this game, if $\mathcal{A}_I$ wants to disguise as a legitimate RN, it must obtain $a_{RN_{i,k}}^X$, which means that it must solve the ECDLP introduced in **Definition 1**. Thus, we have that
$$|Adv_1 - Adv_2| < \frac{O(q_h)^2}{2^l} + O(q_h T_1)$$

**Game 3**: In this game, if $\mathcal{A}_I$ wants to disguise as a legitimate RPM, it must know $a_{RN_{i,k}}^X$ from ciphertext, which also means that it must solve the ECDLP introduced in **Definition 1**. Thus, we have that

$$Adv_3 = Adv_2$$

**Game 4**: In this game, we must consider the false positive caused by the Bloom filter introduced in **Definition 2**. So $\mathcal{A}_I$ can disguise as a legitimate RN with a certain rate according to the size of the Bloom filter. Thus, we have that
$$|Adv_3 - Adv_4| < 2^{-\varepsilon}$$
Thus we have that
$$|Adv_{\mathcal{A}_I}^{RSMS} + Adv_0 - Adv_1 + Adv_1 - Adv_2 + Adv_3 + Adv_3 - Adv_4|$$
$$< |Adv_0| + \frac{O(q_h)^2}{2^l} + O(q_h T_1) + Adv_2 + 2^{-\varepsilon} =>$$
$$|Adv_{\mathcal{A}_I}^{RSMS} + Adv_0 + \_Adv_3 - Adv_4|$$
$$< |Adv_0| + \frac{O(q_h)^2}{2^l} + O(q_h T_1) + Adv_2 + 2^{-\varepsilon} =>$$
$$|Adv_{\mathcal{A}_I}^{RSMS} + Adv_0 - Adv_4|$$
$$< |Adv_0| + \frac{O(q_h)^2}{2^l} + O(q_h T_1) + 2^{-\varepsilon} =>$$
$$|Adv_{\mathcal{A}_I}^{RSMS} - Adv_4| < \frac{O(q_h)^2}{2^l} + O(q_h T_1) + 2^{-\varepsilon}$$
Since $Adv_4 = 0$, we have that
$$Adv_{\mathcal{A}_I}^{RSMS} < \frac{O(q_h)^2}{2^l} + O(q_h T_1) + 2^{-\varepsilon}$$

And we get the result of the final proof of **Theorem 1**. ∎

*Theorem 2.* If, for any polynomial time adversary $\mathcal{A}_{II}$, $Adv_{\mathcal{A}_{II}}^{RSMS}(\kappa) < \frac{O(q_h)^2}{2^l} + \frac{O(q_s + q_e)^2}{2p} + O(q_h T_1) + O(q_h T_2)$, where $q_s$ is a maximum of *Send* queries, $q_e$ is a maximum of *Execute* queries and $q_h$ is a maximum of *Hash* queries, then User accessing Metaverse service resources can be considered as satisfying confidentiality. Note that $\kappa$ is the given security parameter, $T_1$ is the time of solving the ECDLP and $T_2$ is the time of solving the ECDHP.

*Proof.* We also show the proof of **Theorem 2** in a sequence of games following the approach described in [32]. And the success probability of $\mathcal{A}_{II}$ only increases by a negligible amount when moving between the games, as a consequence of *Lemma 1*.

**Game 0**: This game follows the original process, which is called **Game Confidentiality**. Meanwhile, all queries are answered honestly according to the specification of User accessing Metaverse service resources. Thus, we have that
$$Adv_{\mathcal{A}_{II}}^{RSMS}(k) = Adv_0$$

**Game 1**: This game simulates all the oracles for each query, and keeps the lists to store the resulting outputs of the oracles. From the simulation of the games, it is observed that **Game 0** and **Game 1** are indistinguishable. Thus, it holds that
$$Adv_1 = Adv_0$$

**Game 2**: This game proceeds like **Game 0**, but avoids some collisions in the output of hash functions, including $x_i$, and the selection of random values $r$ by a participant *UID*. Thus, **Game 0** and **Game 1** are indistinguishable unless the above collisions occur. And we assume that they are statistically indistinguishable from random strings under random oracle model. Thus, we have that

$$|Adv_1 - Adv_2| < \frac{O(q_h)^2}{2^l} + \frac{O(q_s + q_e)^2}{2p}$$

**Game 3**: In this game, if $\mathcal{A}_{II}$ wants to know the secret value, it must know $f(x_i)$ from $f(x_i) \cdot P$, which means that it must solve the ECDLP introduced in **Definition 1**. Thus, we have that

$$|Adv_2 - Adv_3| < O(q_h T_1)$$

**Game 4**: In this game, if $\mathcal{A}_{II}$ wants to know the secret value, it must know $f(x_j)f(x_i) \cdot P(j \neq i)$ from $f(x_i) \cdot P$ and $f(x_j) \cdot P$, which means that it must solve the ECDHP introduced in **Definition 1**. Thus, we have that

$$|Adv_3 - Adv_4| < O(q_h T_2)$$

Thus we have that

$$|Adv_{\mathcal{A}_{II}}^{RSMS} + Adv_1 + Adv_1 - Adv_2 + Adv_2 - Adv_3 + Adv_3 - Adv_4|$$

$$< 2|Adv_0| + \frac{O(q_h)^2}{2^l} + \frac{O(q_s + q_e)^2}{2p} + O(q_h T_1) + O(q_h T_2) \Rightarrow$$

$$|Adv_{\mathcal{A}_{II}}^{RSMS} + 2Adv_1 - Adv_4|$$

$$< 2|Adv_0| + \frac{O(q_h)^2}{2^l} + \frac{O(q_s + q_e)^2}{2p} + O(q_h T_1) + O(q_h T_2) \Rightarrow$$

$$|Adv_{\mathcal{A}_{II}}^{RSMS} - Adv_4| < \frac{O(q_h)^2}{2^l} + \frac{O(q_s + q_e)^2}{2p} + O(q_h T_1) + O(q_h T_2)$$

Since $Adv_4 = 0$, we have that

$$Adv_{\mathcal{A}_{II}}^{RSMS} < \frac{O(q_h)^2}{2^l} + \frac{O(q_s + q_e)^2}{2p} + O(q_h T_1) + O(q_h T_2)$$

And we get the result of the final proof of **Theorem 2**. ∎

### B. Informal Analysis

**Proposition 1.** RSMS guarantees the continuity of Metaverse service.

**Claim 1. Group Session Key Update** can ensure the continuity of Metaverse service.

Once an RN is considered to be suspicious, the MSRP will allocate one or more new RNs to replace this RN. The MSRP will tell the Metaverse service group and User to authenticate this new RN and update group sesssion key. Then RNs in the Metaverse service group and User will communicate with each other by using the new group session key, which is obtained as described in **Group Session Key Update**.

**Result 1.** *From the claim above it can be derived that RSMS can ensure the continuity of Metaverse service.*

**Proposition 2.** *RSMS preserves the confidentiality of transmitted messages.*

**Claim 2.** The ECDLP is computationally hard by any polynomial-time bounded algorithm.

The asymmetric encryption we use is elliptic curve cryptography. Its security is based on the ECDLP as defined in **Definition 1**. Solving ECDLP is difficult and hasn't efficient algorithm. Therefore, messages sent between the RPM and RN, RN and RN, RN and User are confidential. For the secret share value sent in **Group Session Key Agreement**, it is also protected by elliptic curve point multiplication. So the external malicious attack can't obtain the plaintext of the secret share value and can't be authenticated so that it also can't know the group session key.

**Claim 3.** Only the communicator who knows the symmetric key can decrypt transmitted messages.

The group session key is generated randomly. With **Claim 2**, only the two communicators know the asymmetric key could decrypt the messages. Therefore, group session keys transmitted among the group are computationally infeasible for the adversary to obtain.

During the whole process of RSMS, transmitted messages are always encrypted by applying asymmetric encryption or symmetric encryption among RPMs, RNs, and Users.

**Result 2.** *From the claims above it can be derived that RSMS fulfills confidentiality.*

**Proposition 3.** *RSMS preserves the integrity of transmitted messages.*

**Claim 4.** Only the private key owner can send unique parameters to be confirmed.

In **Initial Authentication**, the RPM can judge whether the message has been tampered with by verifying if $a_{RN_{i,k}}^X{}'$ is in $BF^X$. Then the RN can judge whether the message has been tampered with by verifying $a_{RN_{i,k+1}}^X \cdot P = A_{RN_{i,k+1}}^X{}'$. If true, the message is integrated. In **Re-authentication**, the RPM and RN can judge the integrity of the message in the same way. In **Group Session Key Agreement**, RNs and the User can judge whether the message has been tampered with by verifying $\sum_{i=1}^{N+1} C_i = Q$. Besides, the new session key is verified by computing $s_{serv}^{new}{}' \cdot P = Q + H_5(Q, PK_{pub}) \cdot PK_{pub}$.

Except that, the integrity of $BF^X$ can be guaranteed by verifying the signature of the MSRP.

**Result 3.** *From the claims above it can be derived that RSMS fulfills integrity of transmitted messages.*

**Proposition 4.** *RSMS preserves the non-repudiation of transmitted messages.*

**Claim 5.** Only the private key owner can sign the messages through the secret private key.

Considering that $msk$ is the private key of the MSRP, only the MSRP can generate signature for the public-private key pair, $BF^X$. In mutual authentication phase, only the RPM can decrypt and compute $A_{RN_{i,k}}^X = a_{RN_{i,k}}^X \cdot P$. In **Group Session Key Agreement**, only the User and RNs have their own private key can they decrypt and obtain the secret share value. Then they can compute $f(x_i) \cdot P$ in order to authenticate each other by computing

$$C_i = \left(\prod_{r=1,r\neq i}^{N+1} \frac{-x_r}{x_i - x_r}\right) f(x_i) \cdot P \text{ and verifying } \sum_{i=1}^{N+1} C_i = Q.$$

**Result 4.** *From the claim above, RSMS satisfies the non-repudiation of transmitted messages.*

**Proposition 5.** *RSMS preserves the anonymity.*

**Claim 6.** The output of hash function is indistinguishable from a random string with same length in a statistical sense.

The *PID* of RPM and Users are generated by hash function and $PID^X = H_1(PK_{pub}) \oplus RID^X$. And the initial *PID* of the RN is generated in the same way. When it is re-authenticated, the *PID* will upate by computing $PID_{RN_{i,k+1}}^X = H_6(PID_{RN_{i,k}}^X, a_{RN_{i,k+1}}^X)$.

**Result 5.** *From the claim above, RSMS satisfies anonymity.*

**Proposition 6.** *RSMS preserves the traceability.*

**Claim 7.** The information storage in blockchain is immutable.

The *PID* is stored in blockchain by the RPM. Each time an RN joins a Metaverse service resource pool, the RPM will store its *PID* at present on the blockchain. And if necessary, it can be reovered by the MSRP in order to trace the RN. Due to the information stored in blockchain is immutable, RN's can be traced.

**Result 6.** *From the claim above, RSMS satisfies traceability.*

**Proposition 7.** *RSMS preserves the unlinkability.*

**Claim 6.** The output of hash function is indistinguishable from a random string with same length in a statistical sense.

Each time an RN joins a Metaverse service resource pool, it will update its *PID* by computing $PID_{RN_{i,k+1}}^X = H_6(PID_{RN_{i,k}}^X, a_{RN_{i,k+1}}^X)$. And then the RPM will store its *PID* on the blockchain. Since the output of hash function is indistinguishable from a random string with same length in a statistical sense. So the attacker can't not link the *PID* belongs to the same RN so that it can launch a link attack to obtain the private message of RN.

**Result 7.** *From the claim above, RSMS satisfies unlinkability.*

**Proposition 8.** *RSMS preserves forward and backward security.*

**Claim 2.** The ECDLP is computationally hard by any polynomial-time bounded algorithm.

After the suspicious RN quits the Metaverse service group, the group session key will be updated. The random $s_{serv}^{new}$ is sent to the User and remained RNs in ciphertext by using elliptic curve cryptography. So, others can't know it. And when secret share value is sent to RNs and the User, it is encrypted by their public key using elliptic curve cryptography. So, others also can't know it. Therefore, the suspicious RN can't use the old group session key to decrypt the new message encrypted by the new group session key, which satisfies forward security.

**Claim 8.** Only the private key owner can decrypt the messages encrypted through the corresponding public key can be authenticated.

Before new RNs join the Metaverse service group, the User and RNs will authenticate each other. New RNs must decrypt and obtain the secret share value so that it can be authenticated, which satisfies backward security.

**Result 8.** *From the claims above, RSMS satisfies forward and backward security.*

## VI. Performance and Simulation Analysis

In this section, we analyze the performance of RSMS in terms of computation overhead and communication overhead. Besides, simulation results are presented for evaluating the efficiency of RSMS.

### A. Computation Overhead Analysis

To assess the practical performance of RSMS, we implemented these protocols using Python. In our experimental setup, we leverage the cryptographic tool library known as Miracl Core [31]. Specifically, we opted for the BLS12381 type curve, which provides a robust 128-bit security level. Furthermore, all the commonly used secure hash functions employed in our experiments follow a standardized process. The hash function we use is the SHA256 function. For the execution of our experiments, we utilize a personal computer (PC) running the Ubuntu 20.04.3 operating system. This PC is equipped with an Intel(R)_Core(TM)_i7-10700_CPU@ 2.90GHz and possesses 2GB of memory, providing a suitable environment for conducting our performance evaluations.

TABLE II  MEASUREMENTS OF OPERATIONS AND DELAYS

| Notation | Operations | Time (ms) |
|---|---|---|
| $T_{ECPSM}$ | Elliptic curve point scalar multiplication | 5.64 |
| $T_{MMO}$ | Modular multiplication operation | 0.006 |
| $T_{MIO}$ | Modular inversion operation | 0.007 |
| $T_{HF}$ | Hash function | 0.005 |
| $T_{ECPA}$ | Elliptic curve point addition | 0.027 |
| $T_{MEO}$ | Modular exponentiation operation | 0.058 |
| $T_{BF}$ | Bloom filter verify | 0.024 |

The cryptographic operations used in this paper and their computation overhead are shown in TABLE II.

In **Initial Authentication**, when a RN wants to be authenitcated, it needs $4T_{ECPSM} + 2T_{HF} + 2T_{ECPA}$. When the RPM authenticates the RN, it needs $2T_{ECPSM} + T_{BF} + 2T_{HF} + T_{MIO}$. In **Re-authentication**, when a RN wants to be authenitcated, it also needs $4T_{ECPSM} + 4T_{HF} + 2T_{ECPA}$. When the RPM authenticates the RN, it just needs $2T_{ECPSM} + 2T_{HF} + T_{MIO}$. If with pre-computing, $T_{MIO}$ can be neglected. The XOR operation is also neglected.

In group authentication phase, when parties wants to complete **Group Membership Authentication**, each party needs $2T_{ECPSM} + (N+1)T_{ECPA}$. Then, each party needs $NT_{ECPSM} + T_{HF}$ to complete **Group Session Key Agreement**. In **Group Session Key Update,** each party needs $2T_{ECPSM} + T_{ECPA} + T_{HF}$.

## B. Communication Overhead Analysis

According to the experimental settings, we compute the size of communication packets. In **Initial Authentication**, the RN needs to send 512+160*2=833 bits to the RPM. The RPM needs to send 512 bits to the RN and 1024 bits to the MSRP. In **Re-authentication**, the RN also needs to send 833 bits to the RPM and the RPM needs to send 512 bits to the RN. In addition, the RPM needs to send 1024+160=1184 bits to the MSRP.

In group authentication phase, each party needs to send $N*2*512$ bits to other party totally to complete **Group Membership Authentication**. Then $N*2*512$ bits are also required in **Group Session Key Agreement**. In **Group Session Key Update**, the MSRP needs to send 1024 bits to each party.

TABLE III COMPARISON OF COMMUNICATION COST

| Phase | Entity | Communication cost (in bit) |
|---|---|---|
| Initial Authentication | RN | 833 |
|  | RPM | 1536 |
| Re-authentication | RN | 833 |
|  | RPM | 1696 |
| Group Membership Authentication | Party | $N*1024$ |
| Group Session Key Agreement | Party | $N*1024$ |
| Group Session Key Update | MSRP | $L*1024$ |

Note that $N+1$ is the number of party which panticipates in the group authentication protocol and $L$ is the number of party which isn't suspicious.

## C. Monte Carlo Simulations of Accessing Metaverse service

### C.1) Simulation Settings

In this subsection, we conduct a Monte Carlo simulation experiment to investigate the impact of RSMS on Metaverse service throughput, denoted as *TP*, which represents the number of Metaverse service that can be successfully provided within a specific time period. Our study considers a Metaverse service scenario, where the Metaverse service is already deployed, allowing to access. Within this context, there are several key elements to consider when evaluating *TP*:

- Joining and Leaving Probabilities: We take into account the probabilities associated with RN joining and leaving the Metaverse service group. These dynamics can significantly impact Metaverse service reliability.
- Arriving Probability of Metaverse Service Requests: We consider the likelihood of requesting Metaverse service.
- Timing of Metaverse Service: We analyze the timing of Metaverse service, as this factor can influence how efficiently Metaverse service are delivered.

Through our Monte Carlo simulation, we aim to gain a comprehensive understanding of how RSMS affects *TP*, taking into consideration these various factors and their interplay.

We assume that the probability of RN joining and leaving the Metaverse service group follows exponential distribution with parameters $\lambda_j$ and $\lambda_l$. The probability of Metaverse service request arrival follows Poisson distribution with $\lambda_a$ per day. The time of Metaverse service follows the discrete uniform distribution with parameter $[T_{MS\min}, T_{MS\max}]$. Besides, we set the period of time to 30 mins. The time cost of RSMS we used is showed in VI.A.

### C.2) Simulation Results

We began our investigation by examining how the time of Metaverse service impacts *TP* with RSMS and without RSMS. Initially, We set $\lambda_j = \lambda_l = 0.2$ and $\lambda_a = 0.5$ per day. Then the $[T_{MS\min}, T_{MS\max}]$ is set to different values. The resulting data is presented in Figure 5.

The findings reveal that the introduction of RSMS has a relatively minor impact on *TP* when compared to scenarios without RSMS, as depicted in Fig.5. Notably, it demonstrates that as the time of Metaverse service increases, *TP* experiences a gradual decline, with the values remaining closely aligned. In other words, the implementation of RSMS imposes minimal additional overhead on the system.

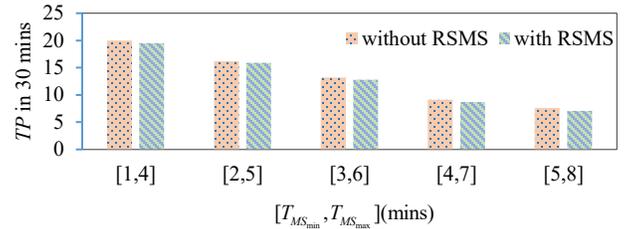

Fig.5. The throughput in different $[T_{MS\min}, T_{MS\max}]$(mins).

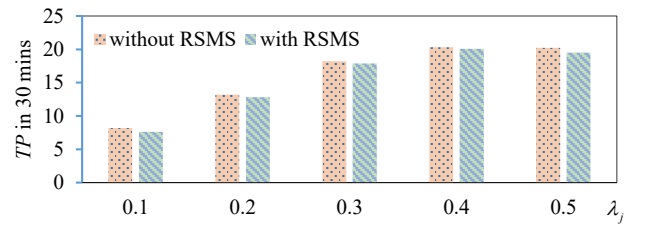

Fig.6. The throughput in different $\lambda_j$.

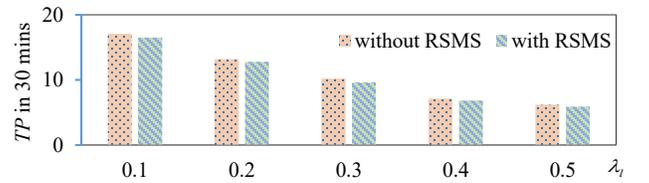

Fig.7. The throughput in different $\lambda_l$.

Besides, we explore the impact of the probability of RN joining and leaving the Metaverse service group on *TP* respectively with RSMS and without RSMS. In the former, we set $\lambda_l = 0.2$ and $\lambda_a = 0.5$ per day. The $[T_{MS\min}, T_{MS\max}]$ is set to [3 mins, 6 mins]. Then $\lambda_j$ is set to different values. In the latter, we fix the $\lambda_j = 0.2$ and $\lambda_l$ is set to different values. And other parameters are the same. Once again, our findings demonstrate that RSMS has a relatively limited impact on *TP* in both cases. Whether it's the probability of RN joining or leaving the Metaverse service group, the system maintains its Metaverse service throughput with minimal fluctuations. And Fig.6 and Fig.7 show the experimental results.

## VII. Conclusion and Future Work

This paper focuses on the provision of a reliable and secure Metaverse service. We achieve this through the design of a a novel mechanism, named *R*eliable and *S*ecure *M*etaverse *S*ervice (RSMS). The design takes into account Metaverse service features and characteristics of entity in Metaverse service system framework. Additionally, security analysis and simulation experiments validate the mechanism's effectiveness.

In the future, we will further explore how to conduct real-time monitoring and security assessment of RNs. Based on this, we can detect suspicious RNs in a more timely manner. At the same time, we will also study some mechanisms to encourage RNs to act honestly to improve the stability of Metaverse service.

**Yanwei Gong** is currently pursuing the Ph.D. Degree in Cyberspace Security at Beijing Key Laboratory of Security and Privacy in Intelligent Transportation, Beijing Jiaotong University, China. His interests include identity authentication protocol related to MEC and fully homomorphic encryption acceleration.

**Xiaolin Chang** is currently a professor at the School of Computer and Information Technology, Beijing Jiaotong University, China. Her current research interests include Edge/Cloud computing, Network security, security and privacy in machine learning. She is a senior member of IEEE.

**Jelena Mišić** is Professor of Computer Science at Ryerson University in Toronto, Ontario, Canada. She has published over 120 papers in archival journals and close to 200 papers at international conferences in the areas of wireless networks, in particular wireless personal area network and wireless sensor network protocols, performance evaluation, and security. She serves on editorial boards of IEEE Transactions on Vehicular Technology, Computer Networks, Ad hoc Networks, Security and Communication

Networks, Ad Hoc & Sensor Wireless Networks, Int. Journal of Sensor Networks, and Int. Journal of Telemedicine and Applications. She is a Fellow of IEEE and Member of ACM.

PLACE PHOTO HERE

**Vojislav B. Mišić** is Professor of Computer Science at Ryerson University in Toronto, Ontario, Canada. He received his PhD in Computer Science from University of Belgrade, Serbia, in 1993. His research interests include performance evaluation of wireless networks and systems and software engineering. He has authored or co-authored six books, 20 book chapters, and over 280 papers in archival journals and at prestigious international conferences. He serves on the editorial boards of IEEE transactions on Cloud Computing, Ad hoc Networks, Peer-to-Peer Networks and Applications, and International Journal of Parallel, Emergent and Distributed Systems. He is a Senior Member of IEEE and member of ACM.

PLACE PHOTO HERE

**Yingying Yao** received PhD degree in cyberspace security from Beijing Jiaotong University, in 2021. She is currently a lecturer in information security department at Beijing Jiaotong University. Her current research interests include applied cryptography, vehicular security and privacy.